\documentclass[12pt,a4paper]{article}
\usepackage[utf8]{inputenc}
\usepackage[T1]{fontenc}
\usepackage{microtype}
\usepackage{lmodern}

\usepackage{amsmath, amssymb, amsthm}
\usepackage{mathtools}
\usepackage{geometry}
\usepackage{graphicx}
\usepackage{hyperref}
\usepackage[capitalise,noabbrev]{cleveref}

\usepackage[round,authoryear,sort]{natbib}
\makeatletter
\newcommand{\textcite}{\@ifnextchar[{\hg@textcite@opt}{\hg@textcite@noopt}}
\def\hg@textcite@opt[#1]#2{\citet[#1]{#2}}
\def\hg@textcite@noopt#1{\citet{#1}}
\newcommand{\parencite}{\@ifnextchar[{\hg@parencite@opt}{\hg@parencite@noopt}}
\def\hg@parencite@opt[#1]#2{\citep[#1]{#2}}
\def\hg@parencite@noopt#1{\citep{#1}}
\makeatother

\renewcommand{\d}{\mathrm{d}}
\newcommand{\F}{\mathcal{A}}
\newcommand{\G}{\mathcal{G}}
\DeclareMathOperator{\Ad}{Ad}
\DeclareMathOperator{\tr}{tr}

\title{Ghosts that Connect}
\author{Henrique Gomes\thanks{University of Bonn and Oriel College, University of Oxford. \texttt{gomes.ha@gmail.com}}}
\date{\today}

\begin{document}
\maketitle

\begin{abstract}
What is a Faddeev--Popov ghost? The textbook answer is that it is a trick: a way of writing a functional determinant as a term in the action, and nothing more. \textcite{Dougherty2021} and \textcite{DoughertyRead2026} have argued that this undersells the ghost, which on their account encodes classical structure of the configuration space, viz.\ the degree-one generators of the Chevalley--Eilenberg complex of the action Lie algebroid. I agree with them that the ghost encodes classical structure. But I think the structure is bigger than they say. The ghost, I argue, is a principal connection $\varpi$ on the bundle $\F \to \F/\G$ of gauge potentials over physical configurations; the BRST operator is the exterior derivative along the orbits; and the Chevalley--Eilenberg ghost coincides with $\varpi$ restricted to the orbits (and every $\varpi$ has the same restriction). The restriction drops the horizontal complement, and I show that gauge-fixed quantisation needs that complement. A gauge-fixing installs it as a pairing of configurations across orbits (a counterpart relation, in Lewis's sense); dressed observables and local counterfactuals rely on this pairing, but it is discarded by the quotient $\F/\G$. In the non-Abelian case Singer's theorem forbids a flat $\varpi$, and there the pairing exists only as parallel transport. Finally, BRST survives gauge-fixing because it is vertical and rigid, and so leaves any such pairing unchanged.
\end{abstract}
\tableofcontents

\section{Introduction}
\label{sec:intro}

By reputation, the Faddeev--Popov ghost is a trick. The story goes like this. A gauge theory's path integral counts every physical configuration once per gauge transformation, i.e.\ infinitely often, and \textcite{FaddeevPopov1967} get round this by integrating over one representative per gauge-equivalence class instead. But the procedure requires two anti-commuting auxiliary fields, the \emph{ghosts}, which appear in the action and circulate in loops, though they never turn up as physical states. Moreover, the gauge-fixed action turns out to have a residual symmetry, rigid and nilpotent, mixing gauge potential and ghosts: the BRST symmetry of \textcite{BecchiRouetStora1976} and of \textcite{Tyutin1975}. On this telling, then, ghosts are bookkeeping. A Grassmann integral happens to let one write a determinant as a local term in the action; that is why the ghosts are there, and outside the gauge-fixed integral they have no standing at all.\footnote{For the textbook derivation see \textcite[\S16.2]{PeskinSchroeder1995} or \textcite[\S15.6]{Weinberg1996}. \textcite[\S2]{Dougherty2021} assembles the philosophical version of the view, which he calls the received view, from \textcite{Weingard1988}, \textcite{Redhead2003} and \textcite[145]{Healey2007}. Weingard's way of putting it is that ghosts are ``just an artifact of our notation''.}

\textcite{Dougherty2021} thinks all this gets the order of explanation backwards. On his view ghosts are not introduced by quantisation. They are already there classically, as coordinates on the configuration space of the theory, or more precisely on its gauge structure; and if one starts from that space, quantising a gauge theory is the standard procedure and nothing more. (The configuration space in question is the action groupoid, rather than the set-theoretic quotient $\F/\G$: \textcite[\S3]{Dougherty2021} objects to the quotient that it forgets which gauge transformation takes a potential to an equivalent one, and he wants that information kept.) \textcite{DoughertyRead2026} have since systematised the proposal, under the heading of \emph{non-positive quantisation}.\footnote{\emph{Non-positive quantisation} is the umbrella term \textcite{DoughertyRead2026} use for quantisation procedures whose state space admits negative-norm vectors. Ghosts are the paradigm case; the indefinite-metric formalisms of Gupta--Bleuler and the like are others. This is a different axis from their monist/pluralist one, which concerns not classical content but how the nominally different quantisation procedures (canonical, path-integral, Dirac's constrained-Hamiltonian method, DeWitt--Faddeev--Popov, BRST, Batalin--Fradkin--Vilkovisky, Batalin--Vilkovisky) are related to one another. Monism, in \textcite{Dougherty2021}'s strong form, says that they are so many ways of computing one quantisation rather than rival quantisations. That thesis is orthogonal to the present paper, which fixes one prescription and asks what classical geometry its ghosts encode.} I shall take the classical-content thesis for granted here, and part company with Dougherty and Read only over \emph{which} classical structure the ghost encodes. Their answer is algebraic: the ghosts are the degree-one generators of the Chevalley--Eilenberg complex of the action Lie algebroid of $\G$ on $\F$.\footnote{The \emph{action Lie algebroid} of $\G$ on $\F$ is the trivial vector bundle $\F \times \mathrm{Lie}\,\G$, with anchor $\lambda \mapsto X_\lambda$, $X_\lambda$ being the orbit-tangent vector field generated by $\lambda \in \mathrm{Lie}\,\G$. Its bracket on sections is the Lie bracket on constant sections, extended by the Leibniz rule, and it is this bracket that encodes the gauge action. The \emph{Chevalley--Eilenberg complex} of the algebroid is $\Lambda^\bullet(\mathrm{Lie}\,\G)^* \otimes C^\infty(\F)$, with a differential that combines bracket and anchor; the cohomology in degree zero is the algebra of infinitesimally gauge-invariant functions on $\F$. See \textcite[\S3]{DoughertyRead2026} for details.} Mine is geometric: the ghost is a principal connection $\varpi$ on the bundle $\F \to \F/\G$ of gauge potentials over physical configurations, and their algebraic ghost is what is left of $\varpi$ when one restricts it to the gauge orbits. Now, the Faddeev--Popov calculus only ever uses that restriction, so on every one of its formulae the two answers agree. The two answers disagree only about what the ghost is a restriction \emph{of}, and this disagreement matters for the questions on which the calculus is silent: what a gauge-fixing gives us over and above a choice of representatives, why a gauge-fixing can never be global, and what a global quantisation of a non-Abelian theory would have to draw from its classical configuration space.

So here is my question: what classical structure on $\F$ does the ghost encode? I shall count as an answer anything that names a structure $S$ such that, first, the BRST transformation laws follow from the defining properties of $S$, so that the ghost's algebra is read off $S$ rather than posited beside it; and, second, among the things that the quotient $\F/\G$ discards, $S$ supplies everything the quantisation makes use of, so that nothing needs adding beside the ghost. The Chevalley--Eilenberg complex satisfies the first condition, as we will see. Whether it satisfies the second depends on whether gauge-fixed quantisation uses more than what the complex has to offer. I will argue that it does.

The quotient $\F/\G$ forgets two things. The first, already mentioned, is which gauge transformation relates a pair of equivalent potentials. The second is which potential on one orbit corresponds to which on another; and it is this second omission that concerns me, because gauge-fixed quantisation compares configurations across orbits all the time. A slice designates a representative on each orbit, and the Faddeev--Popov measure then records how the slice cuts across the orbits. A dressed observable, such as Dirac's gauge-invariant electron, evaluates a field at the representative designated by the slice. And a counterfactual about a local quantity has a truth value only relative to some such designation. In each case a pairing of configurations across orbits is used, and the pairing is provided by a gauge-fixing. In \S\ref{sec:counterparts} I call the pairing a \emph{counterpart relation}, borrowing the term from Lewis, and I work through the Coulomb-gauge case in some detail. Neither the quotient nor the Chevalley--Eilenberg complex carries this pairing: the differential of the complex acts along each orbit separately, and so is blind to any pairing between orbits.\footnote{The thought that the gauge-quotient leaves out structure has precedents: \textcite{Healey2007} on Aharonov--Bohm and the holonomy interpretation; \textcite{Belot2003} on the limits of treating symmetric models as identical. The modal-semantic parallel with counterpart-theoretic readings (\textcite{Kment2012}, \textcite{Roberts2026}) is taken up in \S\ref{sec:counterparts}. \textcite{Gomes2025} develops the gauge-theoretic version, and \textcite{Gomes2026} the case that quantisation needs it.}

Infinitesimally, a counterpart relation is a complement, at each configuration $A$, to the directions tangent to the orbit. Propagating that complement over the orbits by the group action gives a principal connection $\varpi$ on $\F \to \F/\G$.\footnote{The formal object is a principal connection in the sense of Kobayashi--Nomizu, transposed from finite-dimensional bundles on spacetime to the infinite-dimensional bundle of field space. \S\ref{sec:geometry} gives the construction; for the original treatment, \textcite{Singer1978}; for the rigorous modern mathematical treatment, \textcite{DiezRudolph2019}; for the complete bibliography, \textcite{GomesRiello2017}.} \S\ref{sec:geometry} shows that the BRST algebra follows from $\varpi$'s defining properties together with the vertical part of its structure equation. In detail: $sA = D_A\eta$ says that $\varpi$ returns the generator of any vertical vector; $s\eta = -\tfrac{1}{2}[\eta,\eta]$ is the Maurer--Cartan equation along the orbits; and $s^2 = 0$ says that the orbit foliation is involutive. When $\varpi$ is flat, its horizontal distribution integrates to sections, and the connection tells us nothing that a gauge-fixing did not already. But for the non-Abelian cases covered by Singer's theorem \parencite{Singer1978}, namely $M = S^4$ or $S^3$ with $G = \mathrm{SU}(N)$, there is no global section and no flat connection, so whatever structure extends over the whole of $\F$ has curvature. The algebraic reading cannot register this obstruction in the ghost, whereas the connection reading locates it in the holonomy of $\varpi$ and replaces the section-based counterpart relation by parallel transport, which is path-dependent. Parallel transport along $\varpi$ coincides in many respects with Barbour's best matching \parencite{BarbourBertotti1982, Barbour2003, GrybGomes2021}.

One consequence is that there is nothing anomalous about BRST surviving gauge-fixing. BRST acts on $A$ like an infinitesimal gauge transformation, and one can be forgiven for thinking that gauge-fixing has therefore failed to remove the very thing it was brought in to remove. But gauge-fixing has not failed, since BRST is the exterior derivative along the orbits, i.e.\ a single nilpotent differential with a rigid Grassmann parameter, rather than a function's worth of gauge transformations. \S\ref{sec:brst-preserves} shows that verticality and rigidity together mean that BRST leaves unchanged the counterpart relation installed by a gauge-fixing, whereas a field-dependent gauge transformation would destroy it.

The plan is as follows. \S\ref{sec:setup} sets up the machinery: the divergence, gauge-fixing, the ghosts and BRST. \S\ref{sec:rigidity} defines the counterpart relation fixed by a gauge-fixing, shows what uses it, and shows that BRST leaves it intact. \S\ref{sec:geometry} gives the geometry: the ghost as the functional connection, the BRST operator as the leafwise exterior derivative along the orbits, and, via Gribov--Singer, the impossibility of a flat connection. \S\ref{sec:objections} takes up two objections, that the ghost is only the vertical restriction of $\varpi$ and that the choice of $\varpi$ is conventional, and then compares the view with the Vilkovisky--DeWitt programme, on which one $\varpi$ is privileged. \S\ref{sec:conclusion} says how all this bears on sophistication about symmetry \parencite{Dewar2019, Gomes2026}. The bookkeeping for the antighost and the Nakanishi--Lautrup field is relegated to Appendix~\ref{sec:antighost-scope}.\footnote{Anomalies, the Batalin--Vilkovisky extension of BRST, and the role of $\varpi$ at spatial boundaries lie outside the scope of this paper. For anomalies in the geometric BRST framework, see \textcite{Bertlmann2000}. For BV/BFV, see \textcite{BatalinVilkovisky1981, CattaneoMnevReshetikhin2014}. For the functional connection at boundaries,  see \textcite{GomesHopfmullerRiello2019, DonnellyFreidel2016, CarrozzaHohn2022}.}

\section{The path integral, gauge-fixing, and BRST}
\label{sec:setup}

\subsection{The problem and the remedy}
\label{sec:redundancy}
\label{sec:fp}

Let me fix the setting once and for all. We have Yang--Mills theory on a compact Riemannian four-manifold $M$, with a connected, compact, semisimple gauge group $G$ and a principal bundle $P \to M$. Field space $\F$ is the affine space of $G$-connections on $P$. The gauge group $\G$ is the group of vertical bundle automorphisms of $P$, acting on $\F$ by pullback, and $\F/\G$ is the quotient. The Yang--Mills action $S[A] = -\tfrac{1}{2}\!\int_M \tr(F \wedge \star F)$\footnote{With anti-Hermitian generators $\tr$ is negative-definite, and $F \wedge \star F = \tfrac{1}{2}\, F_{\mu\nu}F^{\mu\nu}\,\mathrm{vol}_g$, so the prefactor reproduces the standard $\tfrac{1}{4}\int_M F^a_{\mu\nu}F^{a\,\mu\nu}$.} is $\G$-invariant, and so the naive path integral
\begin{equation}
Z \;=\; \int_{\mathcal{A}} \mathcal{D}A \, e^{-S[A]}
\label{eq:naive-Z}
\end{equation}
counts each physical configuration as many times as there are gauge transformations, i.e.\ with multiplicity $\mathrm{vol}(\G)$, which is infinite.

By itself the infinite multiplicity does no harm, since in normalised amplitudes
\begin{equation}
\langle \mathcal{O} \rangle \;=\; \frac{\int_{\mathcal{A}} \mathcal{D}A \, \mathcal{O}[A] \, e^{-S[A]}}{\int_{\mathcal{A}} \mathcal{D}A \, e^{-S[A]}}
\label{eq:normalised}
\end{equation}
$\mathrm{vol}(\G)$ is an overall factor in both numerator and denominator and simply cancels. The trouble is that the redundancy reappears at the next order, now as a degenerate kernel of the Hessian. Let $A_0$ be a solution and $H[A_0] := \delta^2 S / \delta A^2 \big|_{A_0}$. Then $\G$-invariance implies $H[A_0] \cdot D_{A_0}\lambda = 0$ for every $\mathfrak{g}$-valued function $\lambda$; that is, every orbit-tangent direction is a zero mode, and there is a function's worth of them. Integrate the Gaussian $e^{-\frac{1}{2}\delta A \cdot H \cdot \delta A}$ over these zero modes and out comes $\mathrm{vol}(\G)$ again, this time as an unsuppressed direction: at one loop $\det H = 0$, and the divergence is back in the partition function.
So the propagator, i.e.\ the inverse of $H$, does not exist, and perturbation theory cannot get started. One's first thought is to pass to the physical quotient $\F/\G$, which would cure both problems at a stroke. Unfortunately $\F/\G$ is defined only as a set of equivalence classes. It carries no parametrisation of its own, so any actual computation has to be carried out in $\F$.

\begin{figure}[h]
\centering
\includegraphics[width=0.65\textwidth]{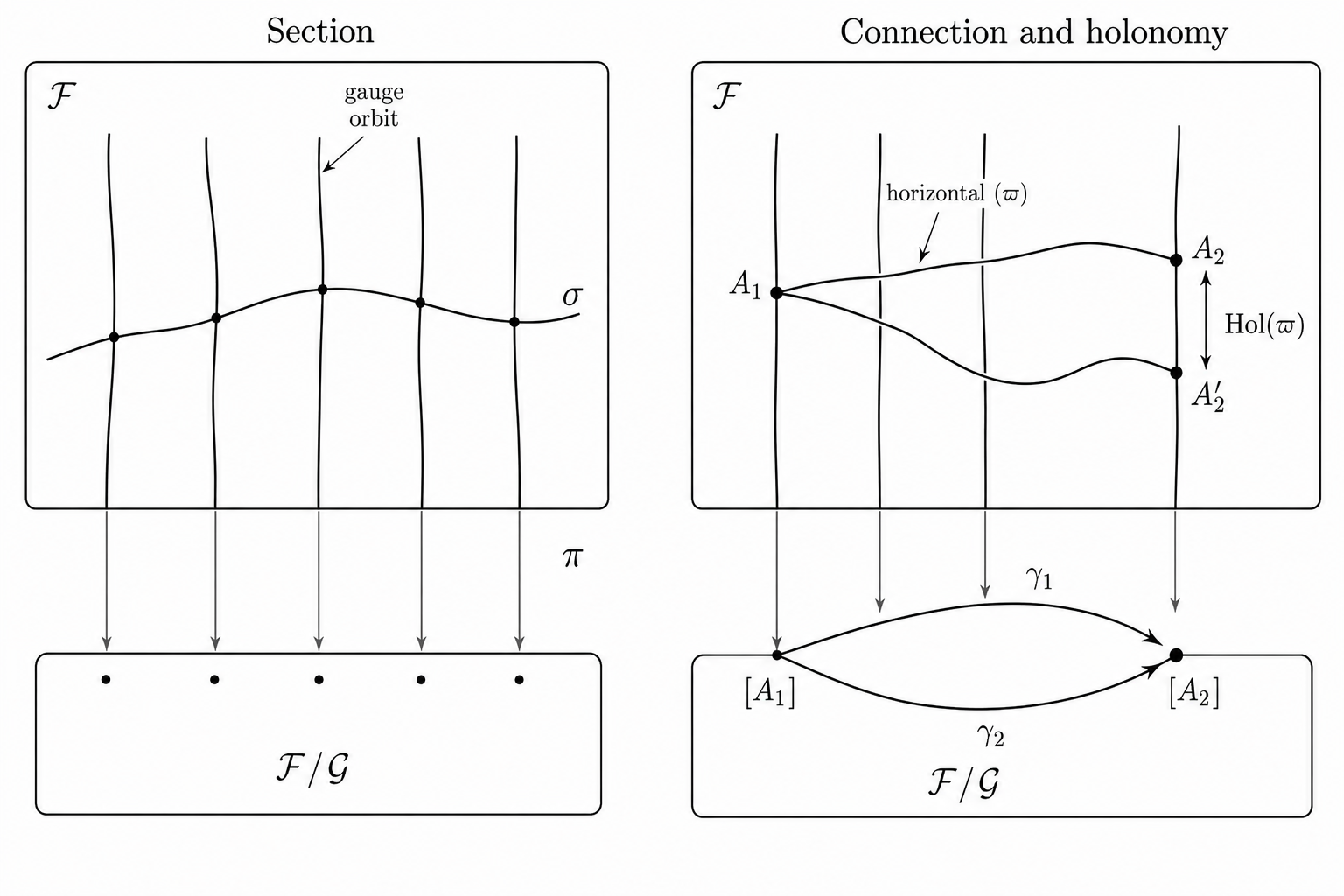}
\caption{Left: a gauge-fixing section $\sigma$ cuts each orbit once, picking one representative per fibre. Configurations $A_1 \in [A_1]$ and $A_2 \in [A_2]$ designated by $\sigma$ are \emph{counterparts relative to $\sigma$}: a between-orbit pairing additional to the within-orbit equivalence individuated by the quotient. Right: in the absence of a global section, alignment is supplied by the horizontal lifts of a connection $\varpi$. Different paths $\gamma_1, \gamma_2 \subset \mathcal{A}/\mathcal{G}$ from $[A_1]$ to $[A_2]$ produce horizontal lifts ending at points $A_2$ and $A_2'$ on the same fibre; their mismatch is the holonomy $\mathrm{Hol}(\varpi)$. The right panel previews \S\ref{sec:geometry}.}
\label{fig:section-counterparts}
\end{figure}

Gauge-fixing is the compromise. We choose a \emph{gauge-fixing functional}, i.e.\ a map $F : \F \to W$ with $W$ a vector space modelled on $\mathrm{Lie}\,\G$, designed to impose one condition per gauge direction; and we work on its zero set $\Sigma_F := F^{-1}(0) \subset \F$, which is a submanifold of field space. On a small enough neighbourhood, where the restriction of $\delta F / \delta A$ to the orbit directions is invertible, $\Sigma_F$ meets each nearby orbit transversally and exactly once.\footnote{The restriction to a region is not a technicality one could remove with more care: no gauge-fixing section extends across the whole bundle. \textcite{Gribov1978} noticed this first, in Coulomb gauge; \S\ref{sec:non-flat} gives the global argument.} There, $\Sigma_F$ serves as a local chart on $\F/\G$: write $A^\sigma \in \Sigma_F$ for the representative on the section and $g \in \G$ for the unique element taking $A^\sigma$ to $A$, and $A \mapsto (A^\sigma, g)$ is a local trivialisation $\F \simeq \Sigma_F \times \G$. Put geometrically, gauge-fixing chooses a complement to the orbit directions (cf.\ \cref{fig:section-counterparts}):
\begin{equation}
T_A \mathcal{A} \;=\; V_A \oplus H_A,
\label{eq:VH-split}
\end{equation}
Here $V_A = \{D_A\lambda : \lambda \in \Omega^0(M, \mathfrak{g})\}$ is the vertical, i.e.\ orbit-tangent, subspace, and $H_A = \ker(\delta F / \delta A)|_A$ is the (integrable) horizontal complement picked out by $F$; at points of $\Sigma_F$ it is the tangent space to the section, and elsewhere the tangent space to the level set $F^{-1}(F[A])$.\footnote{\label{fn:equivariant-extension}One caveat. $\ker(\delta F/\delta A)$ is not in general $\G$-equivariant: in Lorenz gauge, for instance, it is $\{X : \partial^\mu X_\mu = 0\}$ at every $A$, and the adjoint action does not preserve that condition. So it is not itself a principal connection in the sense of \S\ref{sec:geometry}. The connection determined by a gauge-fixing is obtained differently: $\ker(\delta F/\delta A)$ is restricted to $\Sigma_F$ and carried around each orbit by the group action. The result is equivariant by construction, it is flat, and its horizontal leaves are the $\G$-translates of $\Sigma_F$.} The two summands are not on a par: the vertical summand comes with the bundle, whereas the horizontal summand depends on our choice of $F$. In \S\ref{sec:geometry} I will generalise the splitting to cases where no global section exists.

We now change variables from $A$ to $(A^\sigma, g)$. The measure factorises, $\mathcal{D}A = \mathcal{D}A^\sigma \cdot \mathcal{D}g \cdot J[A^\sigma]$, where $J$ is the Jacobian of the decomposition \eqref{eq:VH-split}. Because the action is $\G$-invariant, the $\mathcal{D}g$-integral comes out as $\mathrm{vol}(\G)$, which cancels in normalised amplitudes; what remains is an integral over $\Sigma_F$ weighted by $J$. On a patch where the determinant has a fixed sign (the first Gribov region, where $\mathcal{M}[A]$ is positive, is the standard case) that Jacobian is the determinant of the Faddeev--Popov operator
\begin{equation}
\mathcal{M}[A] \;:=\; \frac{\delta F[A^\lambda]}{\delta \lambda}\bigg|_{\lambda=0},
\label{eq:FP-operator}
\end{equation}
where $A^\lambda$ is the gauge transform of $A$ by $\lambda \in \Omega^0(M, \mathfrak{g})$. In words: $\mathcal{M}[A]$ says how a variation along the orbit through $A$ registers as a variation of the gauge-fixing condition, and $\det \mathcal{M}[A]$ is therefore the Jacobian of the projection from orbit-tangent directions onto $W$, taken at the point where the orbit through $A$ meets the section \parencite{BabelonViallet1979}. To bring a determinant into the exponent, one needs anti-commuting variables. The Berezin identity
\begin{equation}
\det \mathcal{M}[A] \;=\; \int \mathcal{D}\eta \, \mathcal{D}\bar\eta \, \exp\!\left(-\int_M \bar\eta \cdot \mathcal{M}[A] \cdot \eta \right)
\label{eq:berezin}
\end{equation}
turns the Jacobian into a term in the action, and the path integral becomes
\begin{equation}
Z \;=\; \int_{\mathcal{A}} \mathcal{D}A \; \delta(F[A]) \int \mathcal{D}\eta \, \mathcal{D}\bar\eta \; e^{-S[A] - S_{\mathrm{FP}}[A, \eta, \bar\eta]},
\label{eq:Z-fp}
\end{equation}
with $S_{\mathrm{FP}}[A, \eta, \bar\eta] = \int_M \bar\eta \cdot \mathcal{M}[A] \cdot \eta$. The fields $\eta$ and $\bar\eta$ are the \emph{ghost} and \emph{antighost}. On the Faddeev--Popov reading they are nothing but a computational device: not physical fields, not asymptotic states, present only because a Grassmann algebra lets us write the determinant as the functional integral of a local quantity. In the Abelian case $\mathcal{M}[A]=\mathcal{M}$ does not depend on $A$, so the Jacobian factors out along with the volume of the gauge group; in the non-Abelian case it does not, and the ghosts stay. \S\ref{sec:geometry} will come back to these objects and give them a geometric reading.\footnote{The $\delta$-function on $\Sigma_F$ is inconvenient in a perturbative expansion. The standard move is to trade it for a Gaussian peak around $\Sigma_F$, by introducing an auxiliary bosonic field $B$ (the Nakanishi--Lautrup field) and writing the gauge-fixing condition as an exponential weight: 
\begin{equation}
S_{\mathrm{GF}}[A, B] \;=\; \int_M \!\left(B \cdot F[A] - \tfrac{\xi}{2}\, B \cdot B\right).
\label{eq:S-GF}
\end{equation}
The total gauge-fixed action is $\widetilde S = S + S_{\mathrm{FP}} + S_{\mathrm{GF}}$, and the path integral runs unconstrained over $\mathcal{A}$, the ghost fields, and the Nakanishi--Lautrup field.
The Gaussian width $\xi$ is a gauge parameter; physical predictions are independent of it. See \textcite{PeskinSchroeder1995}, \S 16.4. The field $B$ is integrated along the imaginary axis (equivalently $B \mapsto iB$), so that $\int \mathcal{D}B\, e^{-S_{\mathrm{GF}}}$ is a convergent Gaussian yielding the standard quadratic gauge-fixing term $\tfrac{1}{2\xi}\int_M F[A]\cdot F[A]$; the appendix carries the $i$-conventions explicitly.}

\subsection{BRST symmetry}
\label{sec:brst}\label{sec:puzzle}

The total action $\widetilde S$ is no longer gauge-invariant, since $S_{\mathrm{FP}}$ and $S_{\mathrm{GF}}$ both depend on the gauge-fixing functional $F$. It does, however, admit a residual symmetry. With a single anti-commuting parameter $\varepsilon$, define
\begin{equation}
s A_\mu = D_\mu \eta, \qquad s \eta = -\tfrac{1}{2}[\eta, \eta], \qquad s \bar\eta = i B, \qquad s B = 0,
\label{eq:BRST}
\end{equation}
and extend $s$ to products by the graded Leibniz rule, with ghost-number grade $+1$.\footnote{$\varepsilon$ enters multiplicatively: every BRST transformation of a field $\phi$ is $\delta_\varepsilon \phi = \varepsilon\, s\phi$, with $\varepsilon$ suppressed in the displayed conventions \eqref{eq:BRST} and understood throughout. The operator $s$ marks the BRST direction; $\varepsilon$ is the (rigid) parameter scaling the motion.} Then $s\widetilde S = 0$ and $s^2 = 0$ on every field. The operator $s$ is the BRST operator of \textcite{BecchiRouetStora1976} and \textcite{Tyutin1975}.

Three features of \eqref{eq:BRST} matter for what follows. (i) On $A_\mu$, $s$ acts as an infinitesimal gauge transformation with parameter $\eta$; so BRST makes the ghost stand in for a gauge parameter. (ii) On the ghost, $s\eta = -\tfrac{1}{2}[\eta,\eta]$ has the form of a Maurer--Cartan equation.\footnote{For a Lie group $H$ with Lie algebra $\mathfrak{h}$, the canonical left-invariant $\mathfrak{h}$-valued one-form $\theta$ on $H$ satisfies $d\theta + \tfrac{1}{2}[\theta,\theta] = 0$. Why the BRST algebra on the ghost takes this shape is the question \S\ref{sec:ghost-is-varpi} answers.} (iii) The parameter $\varepsilon$ is a single Grassmann scalar. It is not a function on $M$, nor a functional on $\F$: BRST is \emph{rigid}, in the sense that $\varepsilon$ has no spacetime or configuration dependence. The ghost $\eta$, by contrast, is one of the ingredients of the transformation, not its parameter. It is a $\mathfrak{g}$-valued Grassmann field on spacetime, of exactly the type of an anti-commuting gauge parameter, but it is a field of the gauge-fixed theory in its own right. Much of what follows turns on not conflating $\varepsilon$ with $\eta$.

Gauge-fixing does remove the degeneracy it was brought in to remove. Relative to the splitting \eqref{eq:VH-split}, $H[A_0]$ vanishes on $V_{A_0}$ and is generically non-degenerate on $H_{A_0}$;\footnote{`Generically' because at backgrounds with moduli (instantons being the standard case) $H[A_0]$ keeps finitely many zero modes on $H_{A_0}$; these are dealt with by collective coordinates, not by gauge-fixing.} $S_{\mathrm{GF}}$ contributes a quadratic form that is non-degenerate on $V_{A_0}$; put together, they are invertible on the whole of $T_{A_0}\F = V_{A_0} \oplus H_{A_0}$, so the Hessian of $\widetilde S$ exists and so does the propagator. And yet $sA_\mu = D_\mu\eta$ is still there, and BRST acts on $A_\mu$ exactly as an infinitesimal gauge transformation would. Whether this brings the degeneracy back depends on how many such directions survive: a $\lambda(x)$'s worth of such directions would resurrect the original kernel, whereas a single direction would not. \S\ref{sec:brst-preserves} shows that only a single direction survives.

\section{Rigidity and counterparts}
\label{sec:rigidity}

Gauge-fixing does more than cut each orbit once: it fixes a \emph{counterpart relation} across orbits, a between-orbit pairing that is discarded by the quotient $\F/\G$ (\S\ref{sec:counterparts}) and left unchanged by BRST, even though BRST acts on $A$ as a gauge transformation (\S\ref{sec:brst-preserves}).

\subsection{The counterpart relation}
\label{sec:counterparts}

We consider two orbits $[A_1], [A_2] \in \F/\G$. The section picks one representative on each, $\sigma([A_1]) \in [A_1]$ and $\sigma([A_2]) \in [A_2]$, and thereby pairs them, as the canonical representatives relative to $\Sigma_F$. The quotient by itself does nothing of the sort. It knows \emph{that} two configurations are gauge-equivalent but not \emph{how}, i.e.\ not which gauge transformation carries one to the other (as \textcite{Dougherty2021} complains); and about the correspondence between configurations on \emph{different} orbits it knows nothing whatever. A section, on the other hand, answers the question: given a configuration on orbit $[A_1]$, which configuration on orbit $[A_2]$ is its counterpart? Different sections give different answers.

The construction, following \textcite{Gomes2025}, runs as follows. The \emph{dressing} associated with $\Sigma_F$ is the map $g_\sigma : \F \to \G$ sending each configuration $A$ to the unique gauge transformation carrying it to its representative on $\Sigma_F$:
\begin{equation}
F[A^{g_\sigma(A)}] \;=\; 0.
\label{eq:dressing}
\end{equation}
Uniqueness holds within the Gribov region, where the gauge-fixing is non-degenerate. The right action of $\G$ shifts the dressing equivariantly: applying $h \in \G$ to $A$ moves the slice-aligning transformation by $h^{-1}$ on the left,
\begin{equation}
g_\sigma(A^h) \;=\; h^{-1}\, g_\sigma(A),
\label{eq:dressing-equivariance}
\end{equation}
as one reads off uniquely from $F[(A^h)^{h^{-1} g_\sigma(A)}] = F[A^{g_\sigma(A)}] = 0$. Two configurations $A_1$ and $A_2$ are \emph{counterparts relative to $\sigma$} when the same gauge transformation dresses each to its representative on $\Sigma_F$:
\begin{equation}
g_\sigma(A_1) \;=\; g_\sigma(A_2).
\label{eq:counterparts}
\end{equation}
This is an equivalence relation on $\F$. Within an orbit it is trivial: equivariance \eqref{eq:dressing-equivariance} gives $g_\sigma(A^h) = h^{-1} g_\sigma(A)$, so equality of dressings on the same orbit forces $h = \mathrm{id}$. The non-trivial content of counterparthood is therefore strictly between orbits. The associated structural object is the \emph{equilocality relation}\footnote{Written $\varepsilon_\sigma$ in \textcite{Gomes2025}; I write $\mathcal{E}_\sigma$ to keep it apart from the BRST Grassmann parameter $\varepsilon$ of \S\ref{sec:brst}.}
\begin{equation}
\mathcal{E}_\sigma : \mathcal{A} \times \mathcal{A} \to \mathcal{G}, \qquad (A_1, A_2) \mapsto g_\sigma(A_1)^{-1}\, g_\sigma(A_2),
\label{eq:counter}
\end{equation}
which describes the relative dressing of $A_1$ and $A_2$ and is the identity precisely when they are counterparts: $\mathcal{E}_\sigma(A_1, A_2) = \mathrm{id}$ iff $g_\sigma(A_1) = g_\sigma(A_2)$. The same object captures within-orbit relations: for $A_2 = A_1^h$ on the same orbit, equivariance gives $\mathcal{E}_\sigma(A_1, A_1^h) = g_\sigma(A_1)^{-1}\, h^{-1}\, g_\sigma(A_1)$, the dressing-conjugate of $h^{-1}$ --- so $\mathcal{E}_\sigma$ encodes within-orbit gauge equivalence as well as cross-orbit alignment. Equation \eqref{eq:counter} is translation-invariant, since
\begin{equation}
\mathcal{E}_\sigma(A_1^h, A_2^h) \;=\; (h^{-1} g_\sigma(A_1))^{-1}\, h^{-1} g_\sigma(A_2) \;=\; g_\sigma(A_1)^{-1}\, h\, h^{-1}\, g_\sigma(A_2) \;=\; \mathcal{E}_\sigma(A_1, A_2).
\label{eq:counter-equivariance}
\end{equation}
Chain composition is automatic from the group structure:
\begin{equation}
\mathcal{E}_\sigma(A_1, A_2)\, \mathcal{E}_\sigma(A_2, A_3) \;=\; \mathcal{E}_\sigma(A_1, A_3),
\label{eq:counter-properties}
\end{equation}
which ensures that counterparthood threads consistently across triples of orbits. Different gauge-fixings yield different counterpart relations; Coulomb gauge aligns configurations differently from Lorenz gauge, and nothing in $\G$-invariance alone prefers either.

Three constructions use the relation, and I will take them in increasing order of visibility. The first is the Faddeev--Popov integral itself. The change of variables $A \mapsto (A^\sigma, g)$ of \S\ref{sec:fp} is just the counterpart relation written out as coordinates: $A^\sigma$ names the representative that $A$ is paired with, and $g = g_\sigma(A)^{-1}$ the gauge transformation relating them. The Faddeev--Popov operator $\mathcal{M}_A = dF_A \circ \#_A$ (Appendix~\ref{sec:antighost-scope}) is an isomorphism exactly when the orbit direction $V_A$ is transverse to $\ker dF_A$, which at points of the slice is $T_A\Sigma_F$. It degenerates exactly where the splitting $T_A\F = V_A \oplus T_A\Sigma_F$ fails; and $\delta(F[A])\,\det\mathcal{M}_A$ is the measure induced on $\Sigma_F$ by this splitting, from $\mathcal{D}A$ and the Haar measure on $\G$. Now $T_A\Sigma_F$ points from one orbit to the next, so this measure is a between-orbit quantity: the ghost action $\int_M \bar\eta \cdot \mathcal{M}_A \eta$ already encodes where the slice sits relative to the orbits, and not merely the orbit directions themselves. One more observation, which \S\ref{sec:obj-restriction} will use: the counterpart classes are the translates of the slice, since $\{A' : g_\sigma(A') = g_\sigma(A)\}$ is $\Sigma_F^{\,g_\sigma(A)^{-1}}$; and these translates are precisely the horizontal leaves of the flat connection determined by a gauge-fixing (fn.~\ref{fn:equivariant-extension}).\footnote{A caveat: not every quantisation sums over configurations. Locally covariant quantum field theory \parencite{BrunettiFredenhagenVerch2003} does not; it builds observables functorially, assigning an algebra $\mathfrak{A}(M)$ to each admissible background $M$, with injective morphisms coming from isometric embeddings $M \hookrightarrow N$, and there is no path integral in sight. Its covariance is naturality across backgrounds. That is a different relation from a pairing across gauge orbits on a single background, and a more restricted one, since it obtains only where an admissible embedding exists. The pairing that concerns me belongs to the quantisations that do sum over orbits: the Lagrangian path integral and its dressing-based variants.}

The second construction is dressed observables, and here the pairing is used quite explicitly. We take electrodynamics on $\mathbb{R}^3$ in Coulomb gauge, $F[A] = \partial^i A_i$, with the anti-Hermitian conventions of \S\ref{sec:fp}: the Lie algebra of $\mathrm{U}(1)$ is $i\mathbb{R}$, so $A$ is an imaginary-valued one-form, and a gauge transformation $g = e^{\lambda}$, with $\lambda$ an imaginary-valued function, acts by $A \mapsto A^g = A + \d\lambda$ and, on a charged field, by $\psi \mapsto g^{-1}\psi = e^{-\lambda}\psi$.\footnote{The charge is absorbed into $A$. Writing $A = -ieA^{\mathrm{QED}}$ and $\lambda = -ie\lambda^{\mathrm{QED}}$, with $A^{\mathrm{QED}}$ and $\lambda^{\mathrm{QED}}$ real-valued, recovers the textbook form: $A^{\mathrm{QED}} \mapsto A^{\mathrm{QED}} + \d\lambda^{\mathrm{QED}}$, $\psi \mapsto e^{ie\lambda^{\mathrm{QED}}}\psi$, and \eqref{eq:dirac-dressing} becomes $\psi_D = \exp\!\big(-ie\,\Delta^{-1}\partial^i A^{\mathrm{QED}}_i\big)\psi$. Rapid fall-off at infinity is assumed, so that $\Delta^{-1}$ is defined.} The dressing of a potential $A$ is the gauge parameter that carries it to the Coulomb slice, $\lambda_\sigma[A] = -\Delta^{-1}\partial^i A_i$, since $\partial^i(A_i + \partial_i\lambda_\sigma) = 0$; in the notation of \eqref{eq:dressing}, $g_\sigma(A) = e^{\lambda_\sigma[A]}$. Two potentials are Coulomb-counterparts iff $\lambda_\sigma[A_1] = \lambda_\sigma[A_2]$, i.e.\ iff $A_1 - A_2$ is transverse. So the relation says which phase convention at each $x$ in one configuration answers to which in the other; and it says so for configurations on different orbits, which is exactly what the quotient cannot do. \textcite{Dirac1955}'s gauge-invariant electron is the charged field evaluated at the counterpart representative:
\begin{equation}
\psi_D(x) \;:=\; \psi^{g_\sigma(A)}(x) \;=\; g_\sigma(A)^{-1}(x)\,\psi(x) \;=\; \exp\!\big(\Delta^{-1}\partial^i A_i\big)(x)\;\psi(x),
\label{eq:dirac-dressing}
\end{equation}
which is invariant under $A \mapsto A + \d\mu$, $\psi \mapsto e^{-\mu}\psi$, since $\lambda_\sigma$ shifts by $-\mu$. We now consider the gauge-invariant electron propagator $\langle \psi_D(x)\,\bar\psi_D(y)\rangle$. It is an amplitude, summed over configurations on distinct orbits; and each term in the sum is evaluated at whichever representative is picked out by the Coulomb pairing. Changing the pairing changes the amplitude, as it should, since $\psi_D$ with a different dressing is a different operator \parencite{LavelleMcMullan1997}. One last remark on this example. The variation of the dressing along field space, $\delta\lambda_\sigma = -\Delta^{-1}\partial^i\delta A_i$, is (up to sign) the Coulomb-gauge ghost of \S\ref{sec:singer}. The rest of the paper identifies $\eta$ with this variation, i.e.\ with the infinitesimal form of the counterpart relation.

The third construction is counterfactuals about local quantities. One such claim is: had the current been $J'$ rather than $J$, the vector potential at $x$ would have been $A'(x)$. This compares potentials on two orbits at one point, and whether it is true depends on which representatives we compare. Relative to the Coulomb pairing we compare transverse potentials, and the claim has a definite truth value. Relative to no pairing at all, it has no truth value, since a gauge transformation on either side gives $A'(x) - A(x)$ whatever value we like. The same goes for the phase of a charged field at $x$; and, in diffeomorphism-invariant theories, for the value of any field at a spacetime point, where comparing local observables across configurations and gluing regional subsystems into a global state both presuppose an alignment between configurations on different orbits \parencite{Gomes2026}.

The relation is a counterpart relation in Lewis's sense \parencite{Lewis1968, Lewis1986}. In counterpart theory no individual inhabits more than one world. What makes it true that $a$ could have been $\Phi$ is that some individual at another world, $a$'s counterpart there, is $\Phi$; and which individual that is depends on a similarity relation selected by the context. \textcite{Gomes2025} construes the gauge-theoretic relation narrowly, as follows. The individuals are the gauge frames at each point of $M$, i.e.\ the phase conventions of a charged field, or the internal bases of a vector bundle. The worlds are the orbits. A counterpart relation is a group element pairing the frames of one configuration with those of another; and a gauge-fixing supplies one, by \eqref{eq:counter}. Two things follow. First, configurations on the same orbit are related by the gauge transformation between them whatever $\sigma$ we choose (what \eqref{eq:counter} returns is its dressing-conjugate), so the real freedom lies in how configurations on \emph{different} orbits get paired, which is exactly where the counterfactual above lives. Second, the pairing relates representatives and makes no claim of identity across worlds: nothing here requires a gauge frame in one configuration to \emph{be} a frame in another.\footnote{The argument does not require haecceitism in the sense of primitive individuation --- only the claim that \emph{some} non-qualitative pairing is required, whether underwritten metaphysically or semantically.}

Modal semantics has reached, independently, the conclusion that some such pairing is needed once possibilities are individuated qualitatively. \textcite{Kment2012} argues that an anti-haecceitist semantics for chance and counterfactuals cannot make do with bare, qualitatively individuated worlds: it has to enrich them, either with world-descriptions that involve counterpart mappings or with a generalised similarity ordering over individuals, since the qualitative facts do not settle which individual in one world answers to which in another. \textcite{Roberts2026} reaches a parallel conclusion within higher-order modal logic, and pointedly without counterpart theory: ordinary modal discourse, he argues, operates with selector functions that do the job counterpart maps do. In our setting the orbits in $\F/\G$ are the qualitatively individuated possibilities, the dressing is the selector, and the three constructions above mount the physical case for needing one. There are other ways to describe what gauge-fixing does, of course. My reason for preferring this one is that the very same structure, taken infinitesimally, is the ghost (\S\ref{sec:geometry}), and that we then have one structure on our hands rather than two (\S\ref{sec:obj-restriction}).

One might hope to get around the counterpart thesis by quantising with relational observables, or by anchoring everything to asymptotic boundary data, or by passing to Dirac-dressed variables. Each of these is a way of fixing a cross-orbit alignment, not an alternative to one. Dirac dressings are the flat case of a functional connection, as \eqref{eq:dirac-dressing} shows; edge-mode and asymptotic prescriptions fix the alignment through boundary data; and relational-observable constructions dress gauge-variant fields into invariant ones in much the same way.\footnote{For Dirac dressings as the flat case, \textcite{GomesHopfmullerRiello2019}. For edge modes, \textcite{DonnellyFreidel2016, CarrozzaHohn2022}. For dynamical reference frames, \textcite{GoellerHohnKirklin2022}.}

\subsection{BRST preserves counterparts}
\label{sec:field-dep}\label{sec:brst-preserves}

Let us go back to the observation at the end of \S\ref{sec:puzzle}. BRST acts on $A_\mu$ as an infinitesimal gauge transformation would, and a gauge symmetry with a function's worth of parameters would bring back the very kernel of the Hessian that was supposed to be removed by gauge-fixing. Recall what those zero modes of $H[A_0]$ were: \emph{bosonic} tangent vectors $D_{A_0}\lambda$, one for each function $\lambda \in \Omega^0(M, \mathfrak{g})$. The gauge-fixing term changed the quadratic form so that these directions dropped out of the kernel. And now BRST hands us an orbit-tangent direction $sA = D_A\eta$ of exactly the same shape.\footnote{``Rigid'' does not mean ``field-independent'' in any naive sense: $sA = D_A\eta$ depends on $A$, through the covariant derivative, and on the ghost field $\eta$. What is the same at every point of $M$ and at every point of $\F$ is the free Grassmann parameter $\varepsilon$ multiplying $s$. The only thing withheld is a function's worth of parameters.}

Two features of BRST save the counterpart relation. One is \emph{verticality}: $sA = D_A\eta$ lies in the orbit-tangent (vertical) directions, with $\eta$ as the orbit-shift parameter. \S\ref{sec:BRST-geom} will say what sort of object $\eta$ is; for the moment we need only that $sA$ never leaves $V_A$. The other is \emph{rigidity}: $\eta$ is one odd field of the gauge-fixed theory, not a parameter $\lambda[A]$ chosen anew at each configuration, so the same $\eta$ enters the transformation at $A_1$ as at $A_2$, and the BRST flow acts uniformly over $\F$. A word on terminology. In \S\ref{sec:brst} `rigid' meant that $\varepsilon$ is a scalar. I am now stretching it to cover the configuration-independence of $\eta$ as an integration field as well; and the cross-orbit argument below turns on the latter, not the former.

Rigidity preserves counterparts. The dressing equivariance \eqref{eq:dressing-equivariance} with $h = \exp(\varepsilon\eta)$ gives, at first order in the Grassmann parameter $\varepsilon$,
\begin{equation}
s\, g_\sigma(A) \;=\; -\eta\, g_\sigma(A),
\qquad
s\, g_\sigma(A)^{-1} \;=\; g_\sigma(A)^{-1}\, \eta.
\label{eq:s-dressing}
\end{equation}
The graded Leibniz rule applied to \eqref{eq:counter} then yields
\begin{equation}
s\, \mathcal{E}_\sigma(A_1, A_2) \;=\; g_\sigma(A_1)^{-1}\, \eta\, g_\sigma(A_2) \;-\; g_\sigma(A_1)^{-1}\, \eta\, g_\sigma(A_2) \;=\; 0.
\label{eq:s-epsilon}
\end{equation}
So the equilocality relation is BRST-closed, and in particular the counterpart condition $\mathcal{E}_\sigma = \mathrm{id}$ is preserved. Notice what the computation assumed. It applied $s$ to a function of two configurations, and so it took a single $\eta$ to transform both of them; in other words it used the diagonal extension of the BRST action to $\F \times \F$. Rigidity enters exactly here: the cancellation in \eqref{eq:s-epsilon} depends on the same $\eta$ acting on both configurations.\footnote{The argument extends to the path-dependent generalisation of \S\ref{sec:geometry}: if $A_2$ is the endpoint of a horizontal lift of a path through $A_1$, the uniform shift carries $A_2$ to the endpoint of the horizontal lift through $A_1^{\exp(\varepsilon\eta)}$.} Suppose instead that the shift were field-dependent, $A \mapsto A^{\exp(\lambda[A](x))}$, so vertical but not rigid. Then $\lambda$ would take different values at $A_1$ and at $A_2$, and the two terms in \eqref{eq:s-epsilon} would fail to cancel. A shift that was rigid but not vertical would fail in another way, by disturbing the alignment between orbits; so both properties are needed.

How rigid must the parameter be, exactly? Translating by a fixed $f \in \mathrm{Lie}\,\G$ preserves counterparts; this is just \eqref{eq:counter-equivariance} in infinitesimal form. Requiring only that a field-dependent parameter be BRST-closed ($sf = 0$, i.e.\ constant along each orbit) does not suffice, for such an $f$ may still differ from one orbit to the next, hence between $A_1$ and $A_2$, and then \eqref{eq:s-epsilon} breaks down. The only freedom left is to shift the gauge condition by a field-independent gauge parameter, which conjugates $\mathcal{E}_\sigma$ by a constant and leaves the condition $\mathcal{E}_\sigma = \mathrm{id}$ untouched.

Two questions remain, and \S\ref{sec:geometry} will answer both. Why does the BRST operator satisfy $s\eta = -\tfrac{1}{2}[\eta, \eta]$? Rigidity does not explain the Maurer--Cartan form; it says only that the parameter is a scalar. And what becomes of the counterpart relation globally? Within a Gribov region a gauge-fixing provides one; but the Gribov--Singer theorem \parencite[Theorem~7]{Singer1978} rules out any continuous section $\sigma : \F/\G \to \F$ for the bundles at issue, so no global cross-orbit identification of the form \eqref{eq:counterparts} is available. One structure will answer both questions.

\section{Ghosts as the connection on field space}
\label{sec:geometry}

The two questions left over from \S\ref{sec:rigidity} (why the BRST algebra has Maurer--Cartan shape, and what takes over from the counterpart relation when there is no global section) have a common answer, and it is a structure every differential geometer knows: a principal bundle with a connection. The bundle in question is the infinite-dimensional bundle $\F \to \F/\G$ of gauge potentials over physical configurations \parencite{DiezRudolph2019}, not a bundle over spacetime. The ghost is its connection one-form, and the BRST algebra drops out of the properties that define a connection on any principal bundle whatever.

\subsection{Field space as a principal bundle}
\label{sec:pfb}\label{sec:varpi}\label{sec:ghost-is-varpi}

The gauge group $\G$ acts freely and properly on field space $\F$ away from reducible configurations.\footnote{A configuration $A$ is \emph{reducible} when its stabiliser in $\G$ is non-trivial. Reducible configurations are rare: in the Sobolev topologies with which $\F$ is usually equipped they form a meagre set, and the generic stabiliser is trivial \parencite{KondrackiRogulski1983, MitterViallet1981}. Over the reducible locus one has a stratification rather than a bundle, handled by slice theorems \parencite{Ebin1970, IsenbergMarsden1982, DiezRudolph2019}. Throughout, I work on the open dense stratum of irreducible configurations.} On that stratum, $\F$ is a principal $\G$-bundle,
\begin{equation}
\pi : \mathcal{A} \longrightarrow \mathcal{A}/\mathcal{G},
\label{eq:F-bundle}
\end{equation}
with vertical tangent subspace $V_A = \{D_A \lambda : \lambda \in \Omega^0(M, \mathfrak{g})\}$ at each $A$, and with $\G$ acting on the fibres by pullback. Note that the vertical distribution doesn't involve a choice: $V_A$ is just the tangent space to the orbit through $A$. A choice does enter with the \emph{horizontal complement}, i.e.\ with a rule saying, at each $A$, which directions alter only the gauge and which alter the physics.

\begin{figure}[h]
\centering
\includegraphics[width=0.7\textwidth]{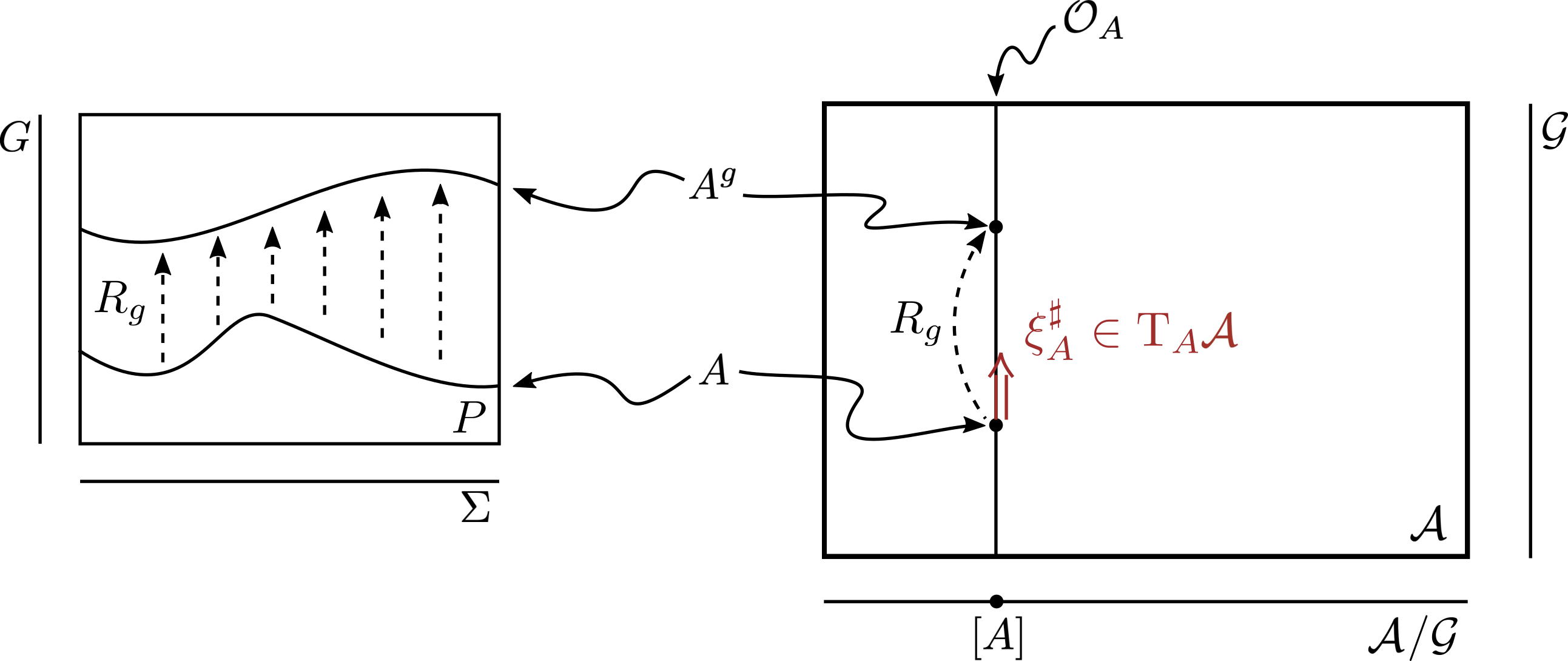}
\caption{The principal bundle of field space (right) by analogy with the finite-dimensional principal bundle $P \to M$ over spacetime (left). The fibre $\mathcal{O}_A$ at $[A] \in \mathcal{A}/\mathcal{G}$ is the gauge orbit through $A$; the structure group $\mathcal{G}$ acts by pullback, with right action $R_g$. A vertical tangent vector $\xi^\#_A \in T_A\mathcal{A}$ is the orbit-tangent vector generated by $\xi \in \mathrm{Lie}\,\mathcal{G}$ via the fundamental-field map $\#_A : \mathrm{Lie}\,\mathcal{G} \to V_A$, $\xi \mapsto \xi^\#_A = D_A \xi$.}
\label{fig:field-space-bundle}
\end{figure}

Recall that on a finite-dimensional principal $G$-bundle $P \to M$ a connection is a $\mathfrak{g}$-valued one-form $\omega \in \Omega^1(P, \mathfrak{g})$ that restricts to the identity on the fundamental vertical fields and is $G$-equivariant under the adjoint action. Its kernel is the horizontal distribution, and its curvature $\Omega = d\omega + \tfrac{1}{2}[\omega, \omega]$ measures how far that distribution is from being integrable. A \emph{functional connection} on $\F$ is the infinite-dimensional analogue \parencite{Singer1978, BabelonViallet1981, GomesRiello2017}: a one-form $\varpi \in \Omega^1(\F, \mathrm{Lie}\,\G)$ such that
\begin{align}
\varpi(X^\#) &= X \qquad \text{for every } X \in \mathrm{Lie}\,\mathcal{G}, \label{eq:varpi-vertical} \\
R_g^* \varpi &= \mathrm{Ad}_{g^{-1}} \varpi \qquad \text{for every } g \in \mathcal{G}. \label{eq:varpi-equivariance}
\end{align}
The kernel of $\varpi$ defines a horizontal distribution $H_A = \ker \varpi|_A$ complementary to $V_A$. This is the splitting \eqref{eq:VH-split} again,
\begin{equation}
T_A \mathcal{A} \;=\; V_A \oplus H_A,
\label{eq:connection-split}
\end{equation}
but now fixed intrinsically by $\varpi$ rather than by a choice of gauge-fixing functional $F$. The vertical summand is the same in both. The horizontal summand here is gauge-equivariant by construction (\eqref{eq:varpi-equivariance}). 

\begin{figure}[h]
\centering
\includegraphics[width=0.4\textwidth]{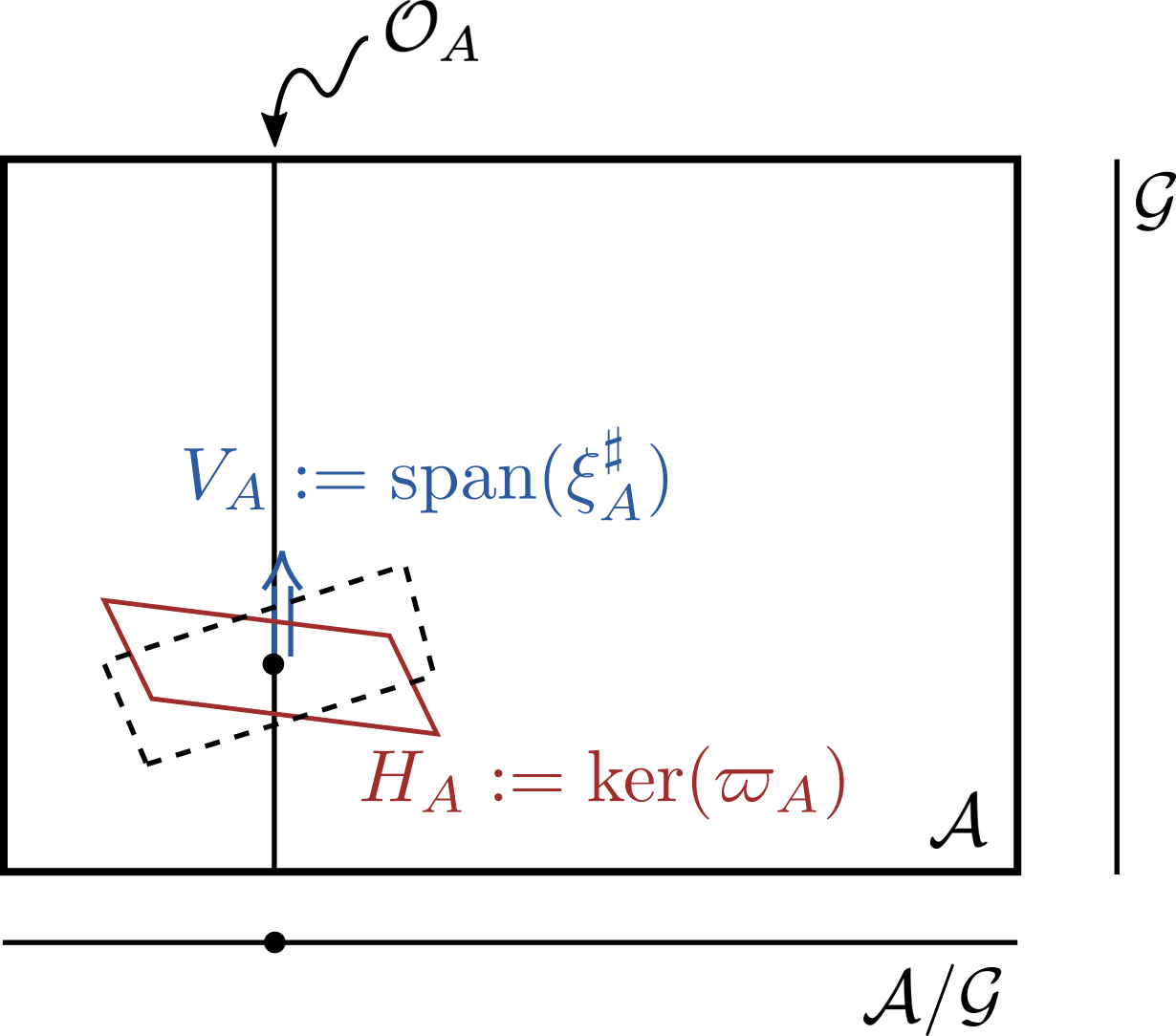}
\caption{A functional connection $\varpi$ specifies a horizontal complement $H_A = \ker(\varpi_A)$ to the vertical subspace $V_A = \mathrm{span}(\xi^\#_A)$ at each $A \in \mathcal{A}$. The horizontal subspace varies with $A$; its variation is encoded in the curvature $\mathfrak{F}$ of $\varpi$.}
\label{fig:VH-split}
\end{figure}

Choosing $\varpi$ amounts to choosing a horizontal complement. The defining equations \eqref{eq:varpi-vertical}--\eqref{eq:varpi-equivariance} belong to the bundle; they say nothing about any gauge-fixing functional $F$. If $\varpi$ happens to be flat then, locally, it makes no difference whether we speak of a section or of a connection. If it is not flat, only the connection is left; and \S\ref{sec:singer} shows that in the non-Abelian Yang--Mills case it cannot be flat.

The curvature of $\varpi$ is
\begin{equation}
\mathfrak{F} \;:=\; \delta \varpi + \tfrac{1}{2}[\varpi, \varpi],
\label{eq:varpi-curvature}
\end{equation}
with $\delta$ the exterior derivative on $\F$ and $[\cdot, \cdot]$ the graded bracket on $\mathrm{Lie}\,\G$-valued forms.\footnote{Some authors use $\mathfrak{F} = \delta \varpi - \tfrac{1}{2}[\varpi, \varpi]$; the sign is absorbed into the bracket convention, and nothing below depends on the choice. I follow \textcite{GomesRiello2017}.} $\mathfrak{F}$ is horizontal by construction: it vanishes whenever any argument is vertical.

Now we can make two identifications. The BRST operator is the leafwise exterior derivative along the gauge orbits,
\begin{equation}
s \;\leftrightarrow\; \delta_V,
\label{eq:s-is-deltaV}
\end{equation}
and the ghost is the connection,
\begin{equation}
\eta \;\leftrightarrow\; \varpi.
\label{eq:eta-is-varpi}
\end{equation}
`Vertical' is being used in two senses here, and I want to keep them apart. In the first, which is intrinsic, $\delta$ is merely evaluated on vertical arguments. The vertical distribution $\mathcal{V} \subset T\F$ belongs to the gauge action ($V_A = \{D_A\lambda\}$ is there with or without a connection), so the fibrewise exterior derivative on $\Gamma(\Lambda^\bullet \mathcal{V}^*)$ is the usual leafwise de Rham differential of the orbit foliation, and needs nothing else. In the second, stronger sense, $\delta$ is split on all of $\Omega^\bullet(\F)$ as $\delta = \delta_V + \delta_H$, with $\delta_H$ acting on the horizontal arguments; that requires a horizontal projector, i.e.\ a connection. The BRST equations \eqref{eq:BRST} need only the first sense, since they evaluate on vertical arguments and nothing else. The second sense enters only when we want $\delta_V$ to act on arguments that are not vertical.

The algebraic reading uses only the first sense. On the irreducible stratum the fundamental-field map $\#_A : \mathrm{Lie}\,\G \to V_A$ is an isomorphism; hence $\Gamma(\Lambda^\bullet \mathcal{V}^*) \cong \Lambda^\bullet(\mathrm{Lie}\,\G)^* \otimes C^\infty(\F)$, and the leafwise de Rham differential is the Chevalley--Eilenberg differential of the action Lie algebroid. What \textcite{Dougherty2021} and \textcite{DoughertyRead2026} call the ghost is the degree-one generator of this complex: the tautological element of $\mathcal{V}^* \otimes \mathrm{Lie}\,\G$ which returns the generator of each vertical vector. By \eqref{eq:varpi-vertical}, that is $\varpi|_{\mathcal{V}}$. And it is $\varpi|_{\mathcal{V}}$ whichever $\varpi$ we take, because on vertical arguments all connections agree. So the algebraic reading agrees with the connection reading restricted to the orbits, and the restriction does not depend on which connection was restricted. The restriction forgets the kernel of $\varpi$, i.e.\ the horizontal complement. Whether the ghost should be identified with the restriction or with the connection is a question that I postpone to \S\ref{sec:obj-restriction}.

The identification \eqref{eq:eta-is-varpi} matches two objects which, on the face of it, live in different places: $\eta(x)$ is a $\mathfrak{g}$-valued Grassmann function on spacetime, $\varpi$ a $\mathrm{Lie}\,\G$-valued one-form on field space. The two are connected through the Lie algebra of the gauge group. $\G$ being the group of vertical bundle automorphisms of $P \to M$, its Lie algebra is
\begin{equation}
\mathrm{Lie}\,\mathcal{G} \;\cong\; \Omega^0(M, \mathfrak{g}).
\label{eq:LieG-iso}
\end{equation}
In other words, an element of $\mathrm{Lie}\,\G$ is a $\mathfrak{g}$-valued field on spacetime, which is just what the ghost $\eta(x)$ is.\footnote{Strictly speaking $\mathrm{Lie}\,\G \cong \Gamma(\mathrm{ad}\,P)$, the sections of the adjoint bundle $\mathrm{ad}\,P = P \times_{\Ad} \mathfrak{g}$, and \eqref{eq:LieG-iso} is its trivialised form, valid globally when $P \to M$ is trivial and only locally otherwise. I write $\Omega^0(M, \mathfrak{g})$ throughout; the reader can substitute $\Gamma(\mathrm{ad}\,P)$ whenever $P$ is non-trivial.} If we evaluate the connection on a tangent vector $\mathbb{X} \in T_A \F$, the result, $\varpi_A(\mathbb{X}) \in \mathrm{Lie}\,\G \cong \Omega^0(M, \mathfrak{g})$, is a $\mathfrak{g}$-valued function on $M$; at each spacetime point $x$ it returns the component of $\mathbb{X}$ that is pure gauge at $x$, and that is the ghost field at $x$. The Grassmann character of $\eta$ also has a geometric source. One-forms on $\F$ anticommute under the wedge product; pulled back under \eqref{eq:eta-is-varpi}, this reads $\eta(x)\,\eta(y) = -\eta(y)\,\eta(x)$, which is to say that $\varpi$, viewed pointwise in spacetime, is a one-form valued in an anticommuting algebra. So $\eta$ as a field and $\varpi$ as a one-form are a single object.

These identifications reproduce the BRST transformation laws of \S\ref{sec:brst}.\footnote{One consequence of the type-clarification: the transposition from differential-form grading on $\mathcal{A}$ to particle-statistics grading on $M$ --- ghosts treated as fermionic for calculational purposes even though they are not fermions --- is a feature of Berezin integration.} For the gauge potential, the vertical derivative gives
\begin{equation}
\delta_V A_\mu \;=\; D_\mu \varpi,
\label{eq:deltaV-A}
\end{equation}
which under \eqref{eq:eta-is-varpi} is $s A_\mu = D_\mu \eta$. On this reading $sA$ is a $\mathrm{Lie}\,\G$-valued one-form on $\F$: evaluated on a tangent vector $\mathbb{X}$, it returns the vertical component of $\mathbb{X}$, realised as a Lie-algebra-valued field on $M$. The familiar picture of a bosonic shift of $A$ along its orbit results from treating $\eta$ as a c-number gauge parameter. Verticality, in this language, says that $sA$ lands in $V_A$ whatever $\mathbb{X} \in T_A\F$ it is evaluated on.\footnote{The derivation proceeds as follows: from the field-dependent gauge transformation $A^\beta = \beta^{-1}(A + \mathrm{d})\beta$, we compute $\delta A^\beta$ and project onto vertical variations to get $\delta_V A^\beta = - D_{A^\beta}(\beta^{-1}\delta_V \beta)$. Identifying $\eta = \varpi$ with $\varpi := \beta^{-1}\delta_V \beta$ reproduces $sA = D_A\eta$ up to a sign reflecting left- versus right-invariant Maurer--Cartan form on $\G$; I fix the sign by \textcite[\S6]{GomesRiello2017}.}

For the ghost, the Cartan structure equation $\delta\varpi = \mathfrak{F} - \tfrac{1}{2}[\varpi, \varpi]$, restricted to vertical arguments on which $\mathfrak{F}$ vanishes, gives\footnote{The bracket $[\varpi, \varpi]$ is itself a vertical 2-form: $[\varpi,\varpi](X,Y) = 2[\varpi(X),\varpi(Y)]$, which vanishes whenever either argument is horizontal, since $\varpi$ does. So $\delta_V\varpi$ and $-\tfrac{1}{2}[\varpi,\varpi]$ are both forms on the orbit foliation, and the equation holds there.}
\begin{equation}
\delta_V \varpi \;=\; -\tfrac{1}{2}[\varpi, \varpi],
\label{eq:MC-vertical}
\end{equation}
which under \eqref{eq:eta-is-varpi} is $s\eta = -\tfrac{1}{2}[\eta,\eta]$. Both BRST laws of \eqref{eq:BRST} fall out of \eqref{eq:s-is-deltaV}--\eqref{eq:eta-is-varpi} together with the defining properties of a connection; neither extra structure nor flatness is required.

\label{sec:BRST-geom}
Nilpotency is automatic. The vertical distribution $\mathcal{V}$ is involutive --- $[X^\#, Y^\#] = [X, Y]^\#$ on fundamental vertical fields --- so the leafwise differential satisfies $\delta_V^2 = 0$. As a consistency check, applying $\delta_V$ to \eqref{eq:MC-vertical} and using the Jacobi identity gives
\begin{equation}
\delta_V^2 \varpi \;=\; -\tfrac{1}{2}\delta_V[\varpi, \varpi] \;=\; -[\delta_V\varpi, \varpi] \;=\; \tfrac{1}{2}[[\varpi, \varpi], \varpi] \;=\; 0,
\label{eq:nilpotency}
\end{equation}
so $s^2\eta = 0$. The same exercise on \eqref{eq:deltaV-A}, using the graded Leibniz rule for $\delta_V$ on $D_A\varpi = \d\varpi + [A, \varpi]$, returns $s^2 A_\mu = 0$. So BRST nilpotency expresses the involutivity of the vertical distribution, whose integral manifolds are the orbits.

\subsection{Why $\varpi$ cannot be flat}
\label{sec:non-flat}\label{sec:singer}
A section and a flat connection are similar, but not quite the same. Clearly, a global section $\sigma$ induces a flat connection: the tangent spaces $T\sigma$, (right-)translated over each orbit by the group action (fn.~\ref{fn:equivariant-extension}), define a horizontal distribution, and the resulting copies of $\sigma$ foliate $\F$ by horizontal leaves. Conversely, via Frobenius' theorem,  a flat connection integrates to such a foliation; and on a \emph{simply connected} base each leaf is itself a global section. So on a simply connected base a global section exists if and only if a flat connection exists.
\footnote{\label{fn:gribov-singer}The original \textcite{Gribov1978} ambiguity in Landau gauge is a different, related failure: $\partial^\mu A_\mu = 0$ intersects most orbits \emph{more} than once. Even the first Gribov region (where $\mathcal{M}_A$ is positive) contains copies; the copy-free \emph{fundamental modular region} is defined by absolute minimisation of $\|A\|^2$ along each orbit and has no closed-form description \parencite{VandersickelZwanziger2012}. The connection-theoretic obstruction $\mathfrak{F} \neq 0$ underlies both failures. \textcite{Singer1978} himself notes that paracompactness of $\F/\G$ allows a partition-of-unity construction using locally-defined slices, but adds that ``this partition of unity argument would be useful if the covering used could be made explicit'' --- a programme that does not appear to have been carried out.}

Three of \textcite{Singer1978}'s results bear directly on the geometric reading. Corollary~4 establishes that for $M = S^4$ (or $S^3$) and $G = \mathrm{SU}(N)$ with $N \geq 2$, no continuous global section $\sigma : \F/\G \to \F$ exists. Theorem~7 strengthens this: the bundle admits no flat connection of any kind, $\varpi_{\mathrm{SdW}}$ included.\footnote{The proof of Theorem~7 uses Theorem~6, which gives the topology of the base: $\F/\G$ is simply connected for $\mathrm{SU}(N)$ with $N > 2$, and $\pi_1(\F/\G) = \mathbb{Z}_2$ for $\mathrm{SU}(2)$ with even Pontryagin index. A flat connection on a simply-connected base would integrate to a section, contradicting Corollary~4. Theorem~7 therefore does independent work only for $\mathrm{SU}(2)$, where a flat connection with non-trivial holonomy need not integrate to a section.} Theorem~9 gives the explicit curvature formula for the Singer--DeWitt connection \eqref{eq:SdW}; for non-Abelian $G$ on $S^4$ that curvature is generically non-zero. Together, these constitute the \emph{Gribov--Singer theorem}: in non-Abelian theories,  gauge-fixing cannot give rise to a continuous global section, nor, equivalently, to a flat connection.\footnote{\textcite{BabelonViallet1981} give the Riemannian-geometric version of the obstruction --- the Gribov horizon as the locus of conjugate points of the horizontal geodesic exponential map. \textcite{MitterViallet1981, GomesRiello2017} treat the bundle structure and standard-form connections in more detail.}

The derivation of \eqref{eq:MC-vertical} used only that $\mathfrak{F}$ vanishes on vertical arguments. The BRST algebra is the vertical projection of the Cartan structure equation, and that projection is insensitive to curvature. Two rival identifications of the ghost therefore agree with the connection reading within a fibre, for different reasons. The identification with the Maurer--Cartan form of $\G$ \parencite{ThierryMieg1980, BonoraCottaRamusino1983} is the flat specialisation of \eqref{eq:eta-is-varpi}: where $\varpi$ is flat on an open $\mathcal{U} \subset \F$ its horizontal distribution integrates (\S\ref{sec:varpi}), giving a local trivialisation $\mathcal{U} \simeq \pi(\mathcal{U}) \times \G$ on which $\varpi$ is the pullback of the Maurer--Cartan form on $\G$. The identification with the Chevalley--Eilenberg generator of the action Lie algebroid \parencite{Dougherty2021, DoughertyRead2026} is a restriction, not a specialisation: it is $\varpi|_{\mathcal{V}}$ (\S\ref{sec:varpi}), the same for flat and curved $\varpi$ alike. So the Maurer--Cartan reading assumes flatness and the algebroid reading ignores it, and neither can register the obstruction described in this section.

\begin{figure}[h]
\centering
\includegraphics[width=0.4\textwidth]{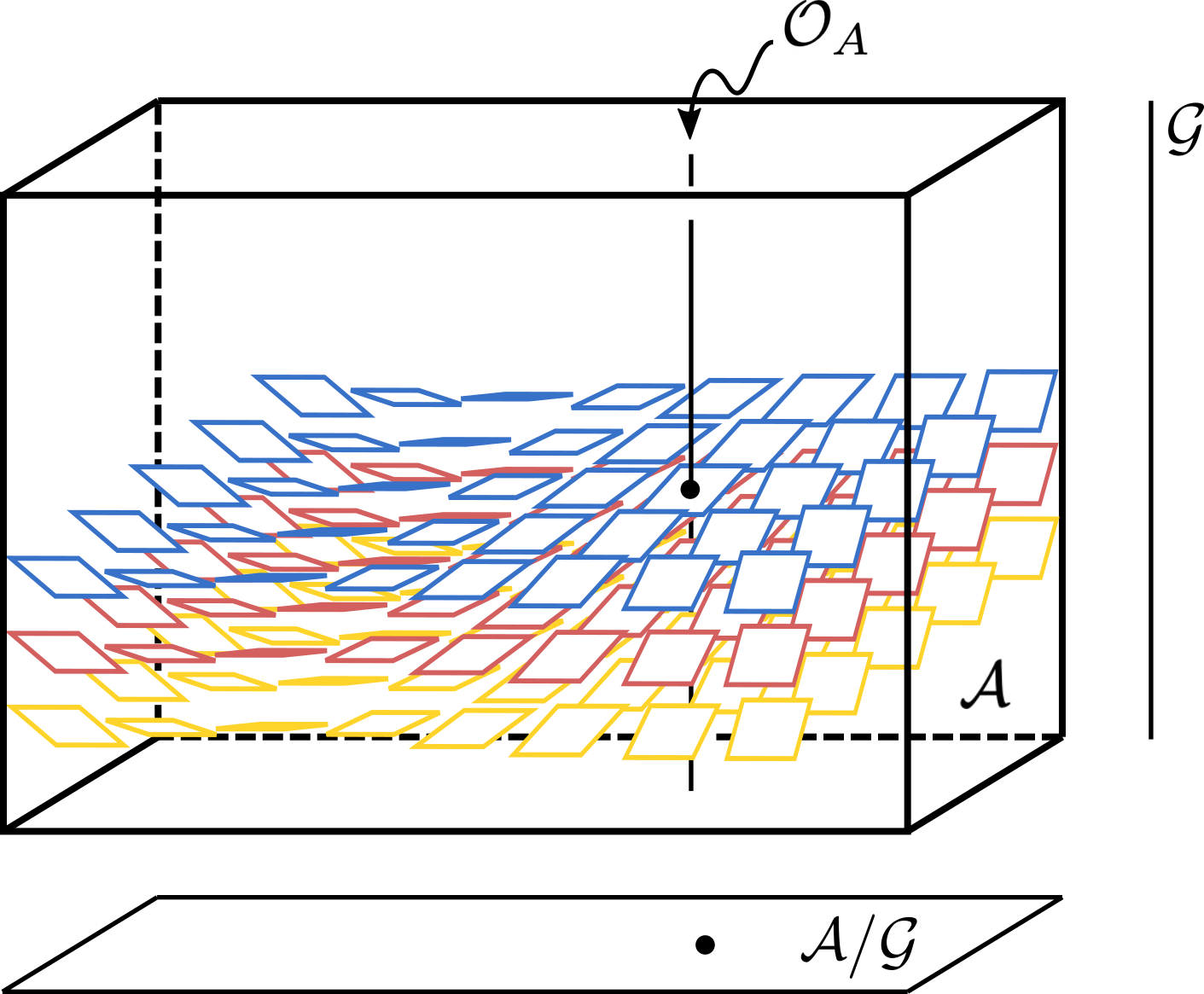}
\caption{A non-integrable horizontal distribution. The horizontal planes (in colour) attached to points along the orbit $\mathcal{O}_A$ do not fit together into a foliation: parallel transport along a closed loop in $\mathcal{A}/\mathcal{G}$ returns the lift to a different point on the same fibre. The mismatch is the holonomy of $\varpi$;  infinitesimally it gives the curvature $\mathfrak{F} \neq 0$.}
\label{fig:anholo}
\end{figure}

The standard explicit example of $\varpi$ is the \emph{Singer--DeWitt connection}, defined by an orthogonality condition in a $\G$-invariant supermetric on $\F$. We fix an ultralocal supermetric --- the spacetime integral of contracted but undifferentiated tangent vectors, $\langle \mathbb{X}, \mathbb{Y}\rangle_A = \int_M \tr(X_\mu Y^\mu)\,\mathrm{vol}_g$, with no derivatives of $\mathbb{X}$ or $\mathbb{Y}$ appearing --- and declare a tangent vector horizontal when it is orthogonal to $V_A$. Using the inner product on the algebra, we define the adjoint operator to $D_A$, namely $D_A^\dagger$. Then a tangent vector $\mathbb{X} \in T_A\F$ decomposes uniquely as $\mathbb{X} = D_A\xi + \mathbb{X}_\perp$ with $D_A^\dagger\mathbb{X}_\perp = 0$, and the Singer--DeWitt connection reads off the vertical coefficient:
\begin{equation}
\varpi_{\mathrm{SdW}}(\mathbb{X}) \;=\; (D_A^\dagger D_A)^{-1} D_A^\dagger \mathbb{X} \;=\; \xi.
\label{eq:SdW}
\end{equation}
For this to make sense $D_A^\dagger D_A$ has to be invertible on $\mathrm{Lie}\,\G$. Its kernel consists of the reducibility parameters, and on the irreducible stratum there are none.

The obstruction is measured by the holonomy of $\varpi$. For a loop $\gamma \subset \F/\G$ bounding a surface $\mathcal{D}$, and a local section $\sigma$ over $\mathcal{D}$,
\begin{equation}
\mathrm{Hol}_\gamma(\varpi) \;=\; \mathcal{P}\exp \oint_\gamma \sigma^*\varpi \;=\; \mathcal{P}\exp \int_{\mathcal{D}} \sigma^*\mathfrak{F} \;\in\; \mathcal{G},
\label{eq:holonomy}
\end{equation}
with the second equality the non-Abelian Stokes theorem \parencite{GomesHopfmullerRiello2019}.

In Abelian gauge theory the bracket on $\mathfrak{g}$ vanishes, so $\mathfrak{F} = \delta\varpi$ for any $\varpi$, and a closed (hence locally exact) one-form $\varpi$ is flat. The Singer--DeWitt construction gives
\begin{equation}\varpi_{\mathrm{SdW}} = \Delta^{-1}\partial^\mu \delta A_\mu= \delta\big(\Delta^{-1}\partial^\mu A_\mu\big),\end{equation}
(since $\Delta$ does not depend on $A$), which is manifestly closed on $\F$; so $\mathfrak{F}[\varpi_{\mathrm{SdW}}] = 0$ identically. Flatness in the Abelian case does not follow from bundle theory in the abstract, since an Abelian connection on a bundle over spacetime can perfectly well be curved. It follows here because, on the field-space bundle, both the supermetric and the group action are field-independent, so the construction returns a field-independent horizontality condition, and the connection it defines is flat. One could build a curved Abelian $\varpi$ by hand, but only by introducing a field-dependent supermetric, i.e.\ a field-dependent horizontality condition.\footnote{For $\mathrm{U}(1)$ on $\mathbb{R}^3$ with rapid-fall-off boundary conditions, $\varpi_{\mathrm{SdW}} = \Delta^{-1}\partial^i\delta A_i$ is the pure-gauge component of $\delta A_i$ at each point. Promoting $\delta A_i$ to a Grassmann-valued field-space one-form, the same formula returns the ghost $\eta(x)$ of Coulomb-gauge QED, and $sA_i = \partial_i\eta$, $s\eta = 0$ are the flat Maurer--Cartan equations of $\mathrm{U}(1)$.} So in the Abelian case the ghost-as-connection collapses, globally, to \textcite{ThierryMieg1980}'s ghost-as-Maurer--Cartan-form, and nothing is lost in passing to the algebroid reading of \textcite{Dougherty2021} and \textcite{DoughertyRead2026}: the horizontal complement is still there, but it is integrable and field-independent, and the vertical restriction plus one global slice gives it back. In the non-Abelian case the collapse fails: the horizontal complement, which carries the curvature, the holonomy and the Gribov obstruction, is invisible to the algebroid, and there is no global slice from which to recover it.\footnote{\textcite{LeinaasOlaussen1981} objected to identifications of ghosts with differential forms on a \emph{finite}-dimensional principal bundle on grounds of nilpotency: products of forms of degree above the manifold's dimension vanish, while products of ghosts do not. Their fix was to take $\G$ as an infinite-dimensional Lie group in its own right. The present framework extends their fix: the ghost lives on the principal bundle of field space, $\F \to \F/\G$, whose structure group is $\G$; this bundle has non-trivial topology as well as infinite-dimensional structure.}

The connection $\varpi$ also supplies a gauge-covariant field-space exterior derivative $\delta_H := \delta - \delta_V$, the field-space analogue of the spacetime covariant derivative $D_\mu = \partial_\mu + [A_\mu, \cdot]$. Quantities built from $\delta_H$ are $\G$-covariant without selecting a section. Where there is no flat limit --- non-Abelian gauge theory globally --- only the covariant apparatus remains definable; see \textcite[\S4]{GomesRiello2017}.

Parallel transport along $\varpi$ coincides in many respects with Barbour's best matching \parencite{BarbourBertotti1982, Barbour2003}. Best matching compares two nearby configurations by sliding one along its orbit until the residual difference is as small as it can be. For the ultralocal supermetric above, the minimising condition on $\mathbb{X} - D_A\xi$ is orthogonality to the orbit, $D_A^\dagger(\mathbb{X} - D_A\xi) = 0$, and that is the horizontality condition of the Singer--DeWitt connection \eqref{eq:SdW} \parencite{GrybGomes2021}. For this connection, horizontal lift is best matching iterated along a path. And the Gribov--Singer theorem tells us how far the procedure can be pushed: best matching between neighbouring configurations is always available, but $\mathfrak{F} \neq 0$ is the measure of how the pairwise matchings fail to knit together into one path-independent, global alignment.

\section{Objections and comparisons}
\label{sec:objections}

The reading I have been developing takes the ghost to be the connection on field space. It has rivals: BRST as Slavnov--Taylor cohomology \parencite{Piguet1995}; the ghost as the Chevalley--Eilenberg generator of the action algebroid \parencite{Dougherty2021, DoughertyRead2026}; and, in the BV/BFV formalism, the ghost as one component of a shifted cotangent bundle, alongside the antifields \parencite{BatalinVilkovisky1981, CattaneoMnevReshetikhin2014}.

The fact that the connection reading can coexist with these rivals is no argument for it. In my view, it is its economy that recommends it. For it adds no degrees of freedom and no extended phase space; rigidity and the Maurer--Cartan equation both follow from the one identification $\eta = \varpi$; and the gauge-fixing sector lives in the horizontal distribution of that same $\varpi$ (\S\ref{sec:obj-restriction}). I now consider three objections. The first is the one the algebraic reading itself suggests: that the ghost is only the vertical restriction of $\varpi$, and the horizontal complement does no work. The second and third concern the choice of $\varpi$, and they pull in opposite directions.

\subsection{The ghost is only the vertical restriction}
\label{sec:obj-restriction}

The objection goes as follows. Every formula of the BRST calculus, whether the transformation laws \eqref{eq:BRST}, nilpotency, or the exactness $S_{\mathrm{GF}} + S_{\mathrm{FP}} = s\Psi_F$ of Appendix~\ref{sec:antighost-scope}, uses $\varpi$ only through its restriction to the orbits, and that restriction is the Chevalley--Eilenberg generator, the same for every $\varpi$ (\S\ref{sec:varpi}). The horizontal complement, then, is surplus: structure that doesn't enter the BRST calculus. The right identification is therefore $\eta = \varpi|_{\mathcal{V}}$, which is the algebraic reading, and to write $\eta = \varpi$ is to credit the ghost with content that lives elsewhere.

I grant the premise: the calculus does use $\varpi$ only through its restriction. It uses the horizontal complement too, but it gets at that through $F$ rather than through $\eta$: the ghost enters the gauge-fixed action as $\int_M \bar\eta \cdot \mathcal{M}_A \eta$ with $\mathcal{M}_A\eta = sF[A]$, which makes the ghost action the Chevalley--Eilenberg differential of the cochain $F$, and $F$ alone is enough to compute it. So the issue concerns bookkeeping: is the slice part of what the ghost encodes, or a separate input standing beside the ghost? For the algebraic reading the slice is a separate input: the complex is canonical, the slice arrives through the gauge-fixing fermion $\Psi_F$, and the transversality of the slice to the orbits has to be imposed afterwards, as a condition of non-degeneracy on $S + s\Psi_F$. This is how \textcite[\S6.2]{DoughertyRead2026} describe BRST quantisation: as picking a non-degenerate representative of the cohomology class of the action. For the connection reading, by contrast, slice and ghost are one object. To see this, we reparametrise $F \mapsto L[A]F$. Factors migrate between $\delta(F)$ and $\det\mathcal{M}_A$, and only the leaf structure of the translates of $\Sigma_F$ survives the reparametrisation; by fn.~\ref{fn:equivariant-extension} and \S\ref{sec:counterparts}, this leaf structure is a flat $\varpi$. Moreover, transversality is built into the very notion of a complement, rather than added later. Both readings, then, require a choice. The connection reading makes it once, as a choice of $\varpi$, and reads off the BRST algebra from the vertical restriction of $\varpi$ and the gauge-fixing sector from its horizontal leaves. This is the economy that I claimed for the connection reading at the start of this section, and locally it settles the comparison.

Globally, however, Singer's theorem separates the two readings. There is no global $\Psi_F$, because there is no global slice; so the algebraic reading has no global gauge-fixing sector to help itself to, and must patch local ones together (fn.~\ref{fn:gribov-singer}). A non-flat $\varpi$, by contrast, does exist globally, and any scheme that quantises across Gribov regions must take its horizontal structure from one. The Vilkovisky--DeWitt programme (\S\ref{sec:obj-counterparts}) is the existing example, though it develops the integral perturbatively about a background. To sum up: the identification $\eta = \varpi|_{\mathcal{V}}$ is the identification $\eta = \varpi$ read inside a fibre. Read on all of $\F$, the identification $\eta = \varpi$ is borne out by how the calculus in fact uses the ghost, and it is the only one of the two identifications whose horizontal structure is globally available.

\subsection{Conventionality}
\label{sec:conventionality}

The counterpart relation depends on a choice of $\varpi$, and nothing in the physics, strictly speaking, privileges one. Coulomb and Lorenz alignments differ; the Singer--DeWitt connection is fixed only once we have fixed a supermetric.

I grant all of this; indeed, it is part of my view. The objection is the inference someone might draw from it: that because the algebroid reading is choice-invariant (the Chevalley--Eilenberg complex is the same whichever $\varpi$ is picked), it already has everything the theory commits us to, and the connection belongs to our representational scheme rather than to the theory.

But although the choice of $\varpi$ is conventional, the need to fix some $\varpi$ is not (\S\ref{sec:obj-restriction}). Charts on a manifold are the obvious comparison: every computation needs one, none is privileged, and none, as a rule, covers the whole manifold. Note also that the connections on $\F \to \F/\G$ form an affine space whose differences are tensorial (horizontal, equivariant, $\mathrm{Lie}\,\G$-valued one-forms), so the gauge-fixing independence of physical amplitudes can be checked as constancy across that space. The algebraic reading records the same invariance in cohomological terms, as the BRST-exactness of the gauge-fixing sector (Appendix~\ref{sec:antighost-scope}); the geometric reading shows, in addition, the space over which the invariance holds.

So I hold that the theory is committed to the class of connections on $\F \to \F/\G$, and so to some counterpart relation, but not to any particular one. A member of the class is fixed on pragmatic grounds, and in the non-Abelian case the member fixed is never flat.

\subsection{The Vilkovisky--DeWitt programme: connection as primitive}
\label{sec:obj-counterparts}

Conventionality can also be met head-on, by privileging a $\varpi$. This is what the \emph{Vilkovisky--DeWitt programme} does \parencite{Vilkovisky1984, DeWitt2003, Pawlowski2003}: it treats $\varpi$ as a primitive of quantisation. A $\G$-invariant supermetric on $\F$, read off the Lagrangian, singles out a preferred $\varpi$. The splitting \eqref{eq:connection-split} then provides a horizontal projector at each $A$, and the geodesic normal fields of the supermetric serve as gauge-invariant coordinates for the fluctuations about a background. One defines the path integral over these horizontal variables, and the resulting effective action is gauge-invariant by construction. The ordinary Faddeev--Popov integral, by contrast, uses $\varpi$ only as the linearised horizontal complement at the background, where section and connection agree to first order. Vilkovisky--DeWitt instead promote $\varpi$ to a global structure, one that defines the very domain of integration.

Were there a preferred supermetric, conventionality would be answered by simply removing the choice. But nothing in this paper requires a preferred $\varpi$, and whether $\varpi$ ought to be promoted to a primitive of quantisation is a question that I leave open.

\section{Conclusion}
\label{sec:conclusion}\label{sec:synthesis}

Let me sum up. I asked what classical structure on $\F$ the ghost encodes. My answer is: a principal connection $\varpi$ on $\F \to \F/\G$. On this answer the BRST laws are the vertical part of $\varpi$'s structure equation, nilpotency expresses the involutivity of the vertical distribution, and the Faddeev--Popov measure on the slice is the measure induced by $\varpi$'s splitting. The Chevalley--Eilenberg reading of \textcite{Dougherty2021} and \textcite{DoughertyRead2026} is a compatible answer, when restricted to the orbits. As far as it goes it is correct. But the restriction is the same for every $\varpi$; so it cannot see what distinguishes one $\varpi$ from another, namely the horizontal complement, and it cannot see what obstructs a flat one. And the horizontal complement \emph{is} used, by gauge-fixed quantisation, by dressed observables and by local counterfactuals, as a pairing of configurations across orbits (\S\ref{sec:counterparts} called it a counterpart relation, and worked it out in Coulomb gauge). The quotient throws that pairing away. Where $\varpi$ is flat, the pairing is given by a gauge-fixing; where Gribov--Singer forbids flatness, it is given by parallel transport along $\varpi$, which is path-dependent, has its holonomy measured by $\mathfrak{F}$, and in many respects coincides with Barbour's best matching (\S\ref{sec:singer}). As for BRST, it survives gauge-fixing because it is the exterior derivative along the orbits; being vertical and rigid, it leaves any such pairing unchanged.

All this bears on \emph{sophistication} about symmetry. \textcite{Dewar2019} formulates sophistication as follows: keep the symmetry-related models, and treat them as representing one possibility, the symmetries being isomorphisms between the models, rather than passing to a reduced theory whose models are the equivalence classes.\footnote{The position has precursors in \textcite{Saunders2003}, \textcite{Belot2003} and \textcite{Caulton2015}. \textcite{Dewar2019} gives the most explicit modern formulation; \textcite{MarchRead2025} give a general statement of the reduction/sophistication distinction, and a fuller bibliography. Dougherty's action groupoid \parencite[\S3]{Dougherty2021} is a sophisticated configuration space in Dewar's sense. Its objects are the potentials on a fixed bundle and its arrows are the gauge transformations between them, so it keeps the symmetry-related models and treats the symmetries as isomorphisms; Dougherty's own gloss, that the groupoid identifies isomorphic potentials without erasing the gauge structure, states the sophisticated view. The identification is mine: Dougherty (his note 15) cites \textcite{Dewar2019} among those who make explicit the desire for the set-theoretic quotient. The Chevalley--Eilenberg complex of the groupoid's algebroid is the infinitesimal algebra of this space.} So sophistication is a thesis about identity conditions for individual models, i.e.\ about when two of them count as one, and it says nothing about how to relate models that are not isomorphic. \textcite{Gomes2026} argues that relations between non-isomorphic models are what the tasks of quantisation require, whether one is summing amplitudes, defining local observables or gluing regions; and that in such settings even a sophisticated theory has to handle its symmetry explicitly. The contribution of the present paper is to locate, within the standard formalism, the point where this happens. The Faddeev--Popov integral makes its cross-orbit pairing explicit in the ghost sector, and BRST leaves that pairing intact; the Chevalley--Eilenberg complex, with its differential acting along each orbit, registers only what is already granted by sophistication, but more is needed for the integral. The connection reading of the ghost says what has to be added: the theory is committed to some pairing of configurations across orbits, though not to any particular one (\S\ref{sec:conventionality}).

There are three questions that I leave open. Non-perturbatively: does the global non-flatness of $\varpi$ show up in anything observable, in confinement, say, or the Gribov horizon, or the non-perturbative measure of the path integral? The Vilkovisky--DeWitt programme is best placed to answer this question \parencite{Vilkovisky1984, DeWitt2003, Pawlowski2003, DonkinPawlowski2012}. Anomalies: the Chevalley--Eilenberg reading locates them in the Lie-algebra cohomology of $\mathcal{G}$, and in the non-flat setting the relevant cohomology should presumably be that of $\varpi$ itself; but so far as I know nobody has worked out the story for $\mathfrak{F} \neq 0$.\footnote{The Dai--Freed treatment of global anomalies is already on the geometric side of this divide. It locates the anomaly in the determinant-line bundle over $\F/\G$, whose Bismut--Freed curvature measures the local anomaly in much the way that $\mathfrak{F}$ measures the obstruction to a flat gauge-fixing \parencite{DaiFreed1994}.} Boundaries: $\varpi$ already does work classically, in the covariant symplectic treatment of bounded regions \parencite{GomesHopfmullerRiello2019, DonnellyFreidel2016, CarrozzaHohn2022}; what its BRST counterpart is, and what $\mathfrak{F}$ does in edge-mode dressings, I do not know.

None of this touches the Faddeev--Popov calculus itself. The Feynman rules stay as they are, and so does the Grassmann integration. The geometric reading supplies an account of the objects the rules are rules \emph{for}, and of why the rules have the form they do.

\subsection*{Acknowledgments}
The author gratefully acknowledges funding from the European Research Council, Grant 101088528
COGY.

\appendix
\section{The ghost and the gauge-fixing doublet}
\label{sec:antighost-scope}

The identifications \eqref{eq:s-is-deltaV}--\eqref{eq:eta-is-varpi} are about the ghost, and about the ghost only. I am not claiming that every Faddeev--Popov auxiliary field is a further piece of the same field-space connection. There are two structures in play, and they should be kept apart: on the one hand the bundle geometry of $\F \to \F/\G$, fixed by the gauge action alone; on the other the implementation of a chosen gauge-fixing functional, which is a later and separate choice. The ghost belongs with the first. The antighost $\bar\eta$ and the Nakanishi--Lautrup field $B$ belong with the second.

The vertical/horizontal split that $\varpi$ supplies at each $A$ reads off vertical motion: $\varpi_A(\mathbb{X}) \in \mathrm{Lie}\,\G$ is the gauge parameter of the vertical part of $\mathbb{X} \in T_A\F$. Choosing $\varpi$ amounts to choosing the horizontal complement; and if $\varpi$ is flat, this is locally the same as choosing a gauge-fixing functional $F$. The difference between the two presentations lies in their global behaviour. The defining equations \eqref{eq:varpi-vertical}--\eqref{eq:varpi-equivariance} belong to the bundle and go on making sense in the cases (\S\ref{sec:singer}) where no $F$ can be chosen globally; the Faddeev--Popov operator, on the other hand, is built directly out of $F$. In Lorenz gauge $W = \Omega^0(M, \mathfrak{g})$ and $F(A) = \partial^\mu A_\mu$; in Coulomb gauge $W = \Omega^0(\mathbb{R}^3, \mathfrak{g})$ and $F(A) = \partial_i A^i$. Either way $\mathcal{M}_A$ is the derivative of $F$ along the vertical directions:
\begin{equation}
\mathcal{M}_A \;=\; dF_A \circ \#_A : \mathrm{Lie}\,\mathcal{G} \to W,
\label{eq:M-as-slice-derivative}
\end{equation}
with $\#_A$ the fundamental-field map of \S\ref{sec:pfb}. $\mathcal{M}_A$ thus measures how the chosen slice $\Sigma_F$ cuts the orbit through $A$, and says nothing about how the bundle itself is put together.

The antighost, for its part, is naturally paired with the gauge-fixing condition rather than with the ghost. For $A \in \Sigma_F$ with $dF_A$ surjective, the dual map $dF_A^* : W^* \to T^*_A \F$ identifies $W^*$ with the conormal directions to $\Sigma_F$, and $\bar\eta_A$ is an odd section of $\Pi W^*$, $\Pi$ denoting parity reversal of the fibres; equivalently, $\bar\eta_A$ is an odd conormal variable for the slice. Its even partner is the Nakanishi--Lautrup field, $B_A \in W^*$. Ghost and antighost are paired through $\mathcal{M}_A$ and through nothing else, via the Berezin identity \eqref{eq:berezin}; without $F$ there is no such pairing.

The clearest way to see how $\mathcal{M}_A$ enters the BRST cohomology is by way of the gauge-fixing fermion, which we define as
\begin{equation}
\Psi_F \;=\; \int_M \bar\eta \cdot \!\left(F[A] - \tfrac{\xi}{2} B\right).
\label{eq:gauge-fixing-fermion}
\end{equation}
Applying the BRST operator via the laws \eqref{eq:BRST} gives
\begin{equation}
s \Psi_F \;=\; \int_M \!\left(i B \cdot F[A] - \tfrac{i\xi}{2} B \cdot B - \bar\eta \cdot \mathcal{M}_A \eta \right).
\label{eq:s-Psi-F}
\end{equation}
The first two terms reproduce $S_{\mathrm{GF}}$ of \eqref{eq:S-GF}, with $B$ replaced by $iB$ (that is the imaginary-axis convention of \S\ref{sec:redundancy}, now made explicit); eliminating $B$ by its algebraic equation of motion gives the usual quadratic gauge-fixing term. The last term is $S_{\mathrm{FP}}$. Hence the whole gauge-fixing sector of $\widetilde S$ is BRST-exact, $S_{\mathrm{GF}} + S_{\mathrm{FP}} = s\Psi_F$, and so cohomologically trivial. Note that $\eta$ and $\bar\eta$ meet only inside this exact expression, and only through $\mathcal{M}_A$. That is the algebraic counterpart of the geometric fact that the antighost belongs to the slice.

The doublet $s\bar\eta = iB$, $sB = 0$ is BRST-contractible, and so contributes nothing to the cohomology. This is why physical amplitudes do not depend on $F$. The same independence of $F$ can be seen geometrically. The curvature of $\varpi$ is intrinsic to $\F \to \F/\G$, whereas $\mathcal{M}_A$ degenerates precisely where the chosen $\Sigma_F$ ceases to cut the orbit transversally, i.e.\ where some non-zero vertical vector $D_A\lambda$ lies tangent to the slice,
\begin{equation}
dF_A(D_A\lambda) \;=\; 0,
\label{eq:gribov-condition}
\end{equation}
so that $\mathcal{M}_A\lambda = 0$. That locus is the Gribov horizon. In the Faddeev--Popov formula it appears as a zero of the ghost--antighost pairing, whereas \S\ref{sec:singer} read it off the global non-flatness of $\varpi$. But the two should not be conflated. The Gribov horizon is a degeneracy of the particular $F$ that happened to be chosen. The curvature of $\varpi$ records an obstruction that no choice of $F$ can remove: the obstruction to replacing connection-based transport by a global flat section.


\end{document}